# Controlling unpredictability in the randomly driven Hénon–Heiles system

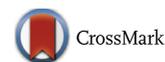

Mattia Coccolo *, Jesús M. Seoane, Miguel A.F. Sanjuán

*Nonlinear Dynamics, Chaos and Complex Systems Group, Departamento de Física, Universidad Rey Juan Carlos, Tulipán s/n, 28933 Móstoles, Madrid, Spain*

**ABSTRACT**

Noisy scattering dynamics in the randomly driven Hénon–Heiles system is investigated in the range of initial energies where the motion is unbounded. In this paper we study, with the help of the exit basins and the escape time distributions, how an external perturbation, be it dissipation or periodic forcing with a random phase, can enhance or mitigate the unpredictability of a system that exhibit chaotic scattering. In fact, if basin boundaries have the Wada property, predictability becomes very complicated, since the basin boundaries start to intermingle, what means that there are points of different basins close to each other. The main responsible of this unpredictability is the external forcing with random phase, while the dissipation can recompose the basin boundaries and turn the system more predictable. Therefore, we do the necessary simulations to find out the values of dissipation and external forcing for which the exit basins present the Wada property. Through these numerical simulations, we show that the presence of the Wada basins have a specific relation with the damping, the forcing amplitude and the energy value. Our approach consists on investigating the dynamics of the system in order to gain knowledge able to control the unpredictability due to the Wada basins.

© 2013 Elsevier B.V. All rights reserved.

## 1. Introduction

There exist a lot of theoretical and experimental works, investigating responses of dynamical systems to external perturbations, such as noise, dissipation or periodic forcing. Depending on the properties of the dynamical systems and the applied perturbation, responses can vary extremely, ranging from practically no effects to a suppressed or an enhanced response [1], regularization of chaotic states [2], chaotification [3], or control of chaotic dynamics [4], among others.

One of the physical phenomenon that exhibits this kind of behavior is the chaotic scattering phenomenon. Chaotic scattering is usually associated with the Hamiltonian equations of motion, that are actually related with chaotic processes. Normally, in this kind of systems, there exists a threshold value of the energy, the escape energy, beyond which the trajectories are unbounded and several exits may appear, therefore the particles are able to leave the scattering region. Since a trajectory might leave the potential well, these systems are usually called *open Hamiltonian systems*. In these cases the particle bounces back and forth in a bounded region, the scattering region, for a certain time before eventually escaping the region towards the infinity.

The phenomenon of chaotic scattering in open Hamiltonian systems has been studied for several years since it has a lot of applications in different fields in science and engineering [5]. Some applications are the analysis of escape from galaxies [6], the study of the interaction between the Earth and the solar wind [7] and many others.

* Corresponding author.
    E-mail address: mattiatommaso.coccolo@urjc.es (M. Coccolo).

On the other hand, in the case of a conservative Hamiltonian system, the total energy is preserved, and thus, it is not possible to talk about attractors nor basins of attraction. A basin of attraction is defined as the set of points that, taken as initial conditions, are attracted to a specific attractor [8]. When we can define two different attractors in a certain region of phase space, two basins exist, which are separated by a basin boundary. This basin boundary can be a smooth curve or can be instead a fractal curve. While we cannot talk about attractors in Hamiltonian systems, we can however define exit basins in an analogous way to the basins of attraction in a dissipative system. In our case, an exit basin is the set of initial conditions that lead to a certain exit.

When boundaries are complicated in a specific region of initial points, a small uncertainty in the position of the initial conditions may yield a greater uncertainty in order to detect the final exit of the trajectory. In fact, there are situations where a small uncertainty in the initial conditions can make them to belong to any of the basins. Nothing can be said because any point on the boundary is arbitrary close to points in all the basins. In the case where we have multiple destinations for the scattering trajectories, the structure of the basins can, eventually, be more complicated [9] and might show the *Wada property* [10]. A basin $B$ verifies the property of Wada if any boundary point also belongs to the boundary of two other basins. In other words, every open neighborhood of a point $x$ belonging to a Wada basin boundary has a nonempty intersection with at least three different basins [11]. Hence, if the initial conditions of a particle are in the vicinity of a Wada basin boundary, we will not be able to be sure by which one of the three exits the trajectory will escape to infinity.

It has been proved that the Wada property can be found in a triangular configuration [11], and typically appears in chaotic scattering systems. Some experimental evidence has been reported in Refs. [12,13] where Wada basins are apparent for higher dimensions. Then, as we said before, the external perturbations can enhance deep modifications in the structures of the basins of the system. In fact, in an open Hamiltonian system, where chaotic scattering phenomena are important, the effects of the dissipation have been an interesting topic [14] because they can induce a new kind of dynamics in the systems [15–18]. Thus, if we add an external perturbation, the topology of the phase space can change abruptly, with the presence of new basins also with the Wada property, as we can see in Ref. [19].

By way of explanations, it is interesting to investigate in which form the influence of an externally driven perturbation and a dissipation term can change the dynamics of a chaotic system. From what we have written above, an external forcing and a damping, can make the system more or less complicated. In other words, the possibility to directly operate on the intensity of an external forcing as a function of the dissipation can reduce the roughness of the basin boundaries until the disappearance of some unpredictabilities, associated to fractal or Wada basins. In the current work, we focus our interest in the numerical analysis of the damped and forced Hénon–Heiles Hamiltonian [20], which is a model of an axisymmetrical galaxy that exhibit chaotic scattering. This is a two dimensional time-independent dynamical system, that shows three different exits for energies greater than the escape energy, so that the system possesses the chaotic scattering phenomenon. In this system we have implemented dissipation and a noisy driven external excitation, in order to study their influence on the topology of the system. To summarize, our goal in this paper is to study the dependence of the Wada basins on the damping and the forcing with a random phase which include the presence of noise [23,22,21]. In other words, we study the possibility to control these unpredictabilities of the system by applying weak external perturbations.

The organization of the paper is as follows. In Section 2 we study the model and the nature of the trajectories. In Sec 3.1 we investigate the external perturbation influence on the unpredictability of the system. In Section 3.2 we investigate the Wada property of the exit basins, changing the bounded excitation and the dissipation at a constant energy and we analyze our data. Finally, a discussion and the main conclusions of this paper are summarized in Section 4.

## 2. Model description

In order to show the influence of an external perturbation on a system with chaotic scattering we take as a prototype model, the Hénon–Heiles Hamiltonian [20], written as

$$H = \frac{1}{2}(\dot{x}^2 + \dot{y}^2) + \frac{1}{2}(x^2 + y^2) + x^2 y - \frac{1}{3}y^3. \tag{1}$$

For energies below the escape energy $E_e = 1/6$, trajectories are bounded and consequently there are no exits. For the energy $E_e = 1/6$, the equipotential line is an equilateral triangle, which is the limit energy at which the motion is bounded as shown in Fig. 1(a). On the other hand, if the energy is larger than this threshold value, the system has three exits with a $2\pi/3$ rotation symmetry, from which the trajectories may escape and go to infinity as shown in Fig. 1(b). Due to the symmetry of the system, the three exits are: exit 1 ($y \to +\infty$), exit 2 ($y \to -\infty, x \to -\infty$) and exit 3 ($y \to -\infty, x \to +\infty$), which are plotted in Fig. 1(b). In this case, there exist three orbits $L_i (i = 1, 2, 3)$, known as Lyapunov orbits, one corresponding to each exit, acting as frontiers: any trajectory that crosses them with an outward-oriented velocity must go to infinity and never come back. We focus our study in a situation with escapes from the scattering region, so from now on we use values of $E > E_e$. We study the Hénon–Heiles system subjected to a bounded noisy excitation (a periodic forcing with a random phase) [19] and a dissipation proportional to the velocity [17]. The equations of motion can be written as

$$\ddot{x} + x + 2xy + \alpha\dot{x} = 0 \tag{2}$$
$$\ddot{y} + y + x^2 - y^2 + \beta\dot{y} = f\cos[\Omega t + \sigma B(t) + \gamma], \tag{3}$$

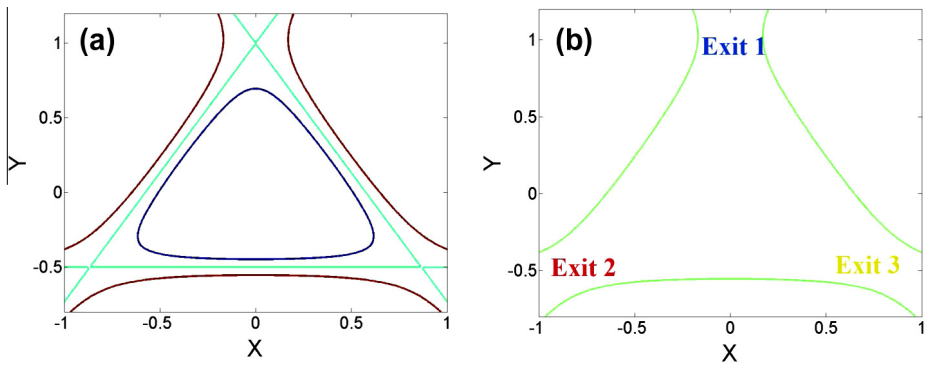

**Fig. 1.** (a) This figure represents the isopotential curves of the Hénon–Heiles system for different values of the energy, in which both bounded and unbounded motions can take place. (b) Plot of the isopotential curve for the unbounded case for energy value $E = 0.21$.

where $\alpha$ and $\beta$ are damping coefficients, $f$ and $\Omega$ are the amplitude and frequency of the external excitation, respectively, $B(t)$ is the standard Wiener process with the amplitude $\sigma$, and $\gamma$ is a random variable uniformly distributed in the interval $[0, 2\pi)$. When $\alpha = \beta = f = 0$ we can recognize the Hénon–Heiles conservative system. From now on, and without any loss of generality, we take $\alpha = \beta = \mu$ as dissipative parameter.

We are studying a two-dimensional time-independent Hamiltonian, so the phase space depends on $(x, y, \dot{x}, \dot{y})$ and one conserved quantity, the energy $E$. Throughout this paper, we will use a Poincaré surface of section to show our results. For that purpose, our choice is $(x = 0, y, \dot{y})$. Thus, the dynamical description of the system can be reduced to a study of the $(y, \dot{y})$ surface. Naturally, the equation of the initial velocity, generically expressed by

$$v_i = \sqrt{\dot{x}^2 + \dot{y}^2} = \sqrt{2E - x_i^2 - y_i^2 - 2x_i^2 y_i + 2/3 y_i^3} \qquad (4)$$

becomes $v_i = \sqrt{2E - y_i^2 - 2/3 y_i^3}$. This simplification is only valid for the initial time $t = 0$, when the dissipation and the external forcing are not yet acting.

In order to study the phase space structure for the Hénon–Heiles Hamiltonian, we compute the exit basins. For that purpose, we compute each trajectory for a large number of initial conditions, $(y, \theta)$, where $\theta$, the *shooting angle*, is the initial angle between the $y$ axis and the trajectory, as shown in Fig. 2. In this way, we can start the trajectories on all the points of the Poincaré section, $x = 0$, and calculate the exit through which every trajectory leaves the potential well. Therefore, knowing the initial conditions related to every trajectory, we color them in a different way, according to the exit through which the trajectories leave the potential, as shown in Fig. 3(a) and (b). We calculate the trajectories and compute the exit basins by using a *symplectic integrator* (SI), that can be mathematically defined as a numerical integration scheme for a specific group of differential equations related to classical mechanics and symplectic geometry. Symplectic integrators form the subclass of geometric integrators that, by definition, are canonical transformations. They are widely used in molecular dynamics, finite element methods, accelerator physics, and celestial mechanics. The trajectories of each particle is followed by numerically solving the Hamiltonian equations from the C-C Algorithm, which is a fourth-order forward symplectic algorithm proposed recently by Chin and Chen [24]. This algorithm can follow the true dynamics longer because it can preserve the symplectic structures of the Hamiltonian equations.

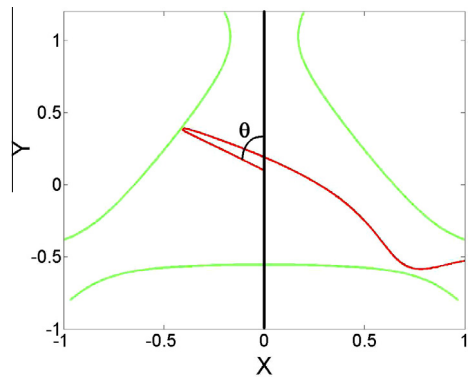

**Fig. 2.** This figure represents the shooting angle $\theta$ of a typical trajectory inside the Hénon–Heiles potential.

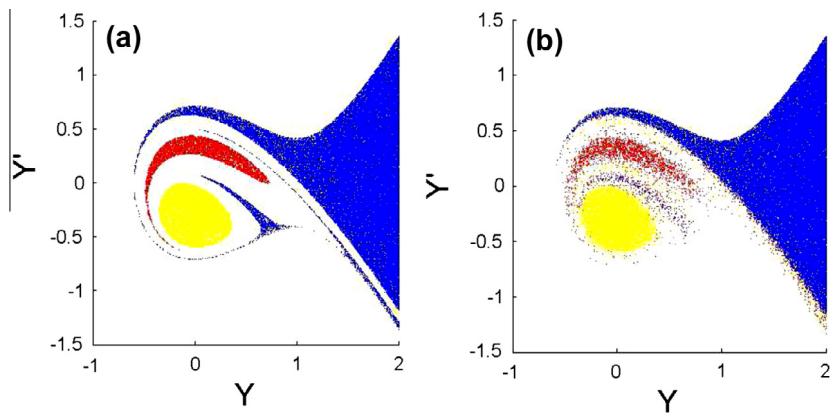

**Fig. 3.** (a) The figure represents the exit basins for dissipative parameter value $\mu = 0.06$, without forcing. (b) The exit basins with both damping, $\mu = 0.06$, and forcing, $f = 0.008$ are plotted. Both figures show the exit basins and each color denotes the exit through which trajectories with that initial condition escape: exit 1 (blue, $(y \to +\infty)$), exit two (red, $(y \to -\infty, x \to -\infty)$) and exit 3, (yellow $(y \to -\infty, x \to +\infty)$). White color inside the color structure denotes the points that do not leave from the scattering region. (For interpretation of the references to color in this figure legend, the reader is referred to the web version of this article.)

As we said earlier in this section, we integrate within the variables of the position, **q**, and the momentum, **p**, as we can see in Fig. 4(a) and (b), including the noisy part of the external excitation,

$$\sigma B(t) = \sqrt{-4D \log(r_1) \sin(2\pi r_2)}, \tag{5}$$

where $\sigma = \sqrt{4D}$ and $B(t) = \sqrt{-\log(r_1) \sin(2\pi r_2)}$ of Eq. (3), $r_1$ and $r_2$ are random numbers in the interval $(0, 1)$. It is possible to appreciate in Fig. 4(a), a trajectory affected only by dissipation, while in Fig. 4(b) the trajectory is affected also by the noisy excitation. Both trajectories start in the same initial condition and are affected by the same amount of dissipation, the only change we introduced is the amount of noise in the random phase of the external forcing.

## 3. Numerical results

In this Section, we are going to provide numerical evidence on the effects of the external perturbation in both the dynamics and the topology of the system and how we can tame the effects of both perturbations in order to reduce the unpredictability of the randomly driven Hénon–Heiles system.

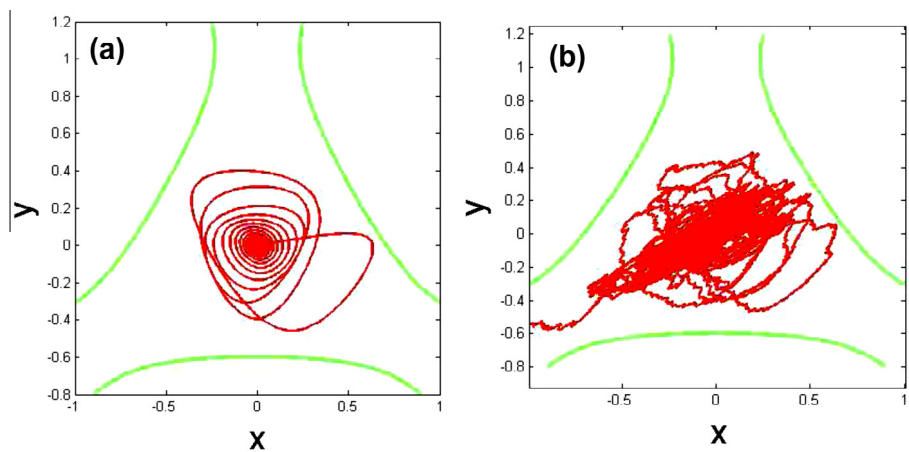

**Fig. 4.** (a) This figure represents a trajectory with dissipation, $\mu = 0.07$ and energy $E = 0.25$ without forcing, with initial conditions $(x_0, y_0) = (0, 0)$ and shooting angle, so called the initial angle between the $y$ axis and the trajectory, $\theta = 0.45\pi$. (b) A trajectory with both damping, $\mu = 0.07$, and forcing amplitude $f = 0.045$, and the same initial conditions as in Fig. 4(a) is plotted. We easily observe the effects of the noisy excitation since this trajectory is similar to a random walk escaping the particle through exit three after a long time.

### 3.1. The effect of the external perturbations on the dynamics of the system

One of the main consequences of chaotic scattering in the Hénon–Heiles system is that, a trajectory may spend a long-time wandering in the vicinity of the scattering region before escaping to infinity from one of the three exits. The transient chaotic dynamics inside the scattering region is governed by the nonattracting chaotic set, also known as the chaotic saddle. This set can be computed through the intersection of the stable manifold that contains the trajectories that will never escape for $t \rightarrow +\infty$, and the unstable manifold that contains the ones that will never escape for $t \rightarrow -\infty$. Both of these sets have singularities, therefore their dimension is fractal [18]. Due to the sensitivity to the initial conditions, characteristic of the chaotic systems, particles can exhibit dramatically different asymptotic behavior. Moreover, if we include dissipation and an external forcing to the system, the exit basins can also change drastically. On the other hand, the phase space might be mixed with KAM islands and chaotic seas, and including a small amount of dissipation can convert the elliptic points inside the islands into sinks, or attractors [15,14]. These dissipation-induced basins of attraction can be intermingled in complicated ways as well, leading to unpredictability or a well defined final state, depending on the initial condition and on the external forcing applied. If we vary the values of the damping and the forcing, the exit basins and the dissipation-induced basins of attractions can show different levels of unpredictability. Basically, it can become difficult to define the exit by which a particle would leave the scattering region, given a set of initial conditions. When the dissipation is high enough and the forcing is low enough, the basin boundaries are smooth. Topologically, this means that the basins are connected and compact. On the other hand, when the external excitation grows up, the basin boundaries become rough and the basins start to mix until they loose the connectedness and compactness. When this happens the boundaries start showing the Wada property and as a consequence the unpredictability in the evolution of the system increases. On the other hand, we focus our research on the unpredictability in a scattering problem in presence of a noisy excitation and dissipation. Here, we investigate the relation between damping, forcing and their effects on the unpredictability in the Hénon–Heiles system. This study is carried out for different values of the energy always beyond the critical energy $E_e = 0.16$ which separates bounded and unbounded trajectories. Therefore, we want to analyze the control of the unpredictability due to forcing and noise, through the energy dissipation. This means that we need to show where the Wada basins appear in function of the energy, the dissipation and the forcing. Naturally, in order to study the above relations, it is important to understand the role of the random phase of the external excitation. We have decided to perturb our system, Eq. (2), with a bounded noisy excitation, that is, a periodic force with a random phase. The value D, in Eq. (5), slightly affects the trajectory with a small amplitude oscillation that depends on the amplitude of the forcing f, and the fluctuation of the function $\cos[\Omega t + \sigma B(t) + \gamma]$, as we can see in Fig. 5(a)–(d). In these figures, in fact, we show how both the forcing and the noisy excitation act on the system. Fig. 5(a) and (c) show the intensity of the noise every iteration. The difference between the figures is the amplitude of the noise D and its effects on the intensity of the signal. Actually, it is possible to appreciate in the graphs the difference of magnitude in the scales. The other two figures (b) and (d) show the intensity of the external forcing, with the same amplitude f, but with the two previous noise signals inside. In other words, those figures show a typical effect of the bounded noise. Its use assures us that, even if integrated along with $(x,y)$, it never overcomes the trajectories of the system but only affects them as a bounded perturbation.

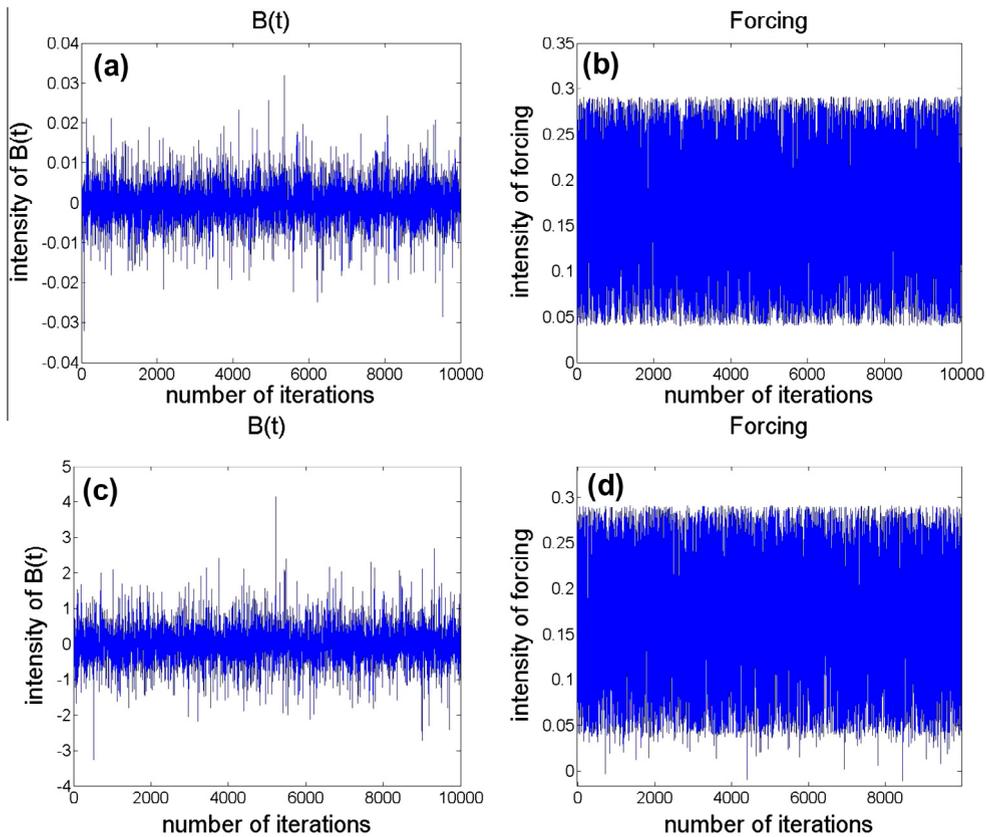

**Fig. 5.** Figures (a) and (c) show the intensity of the noise, as shown in Eq. (5), $\sigma B(t) = \sqrt{-4D \log(r_1)} \sin(2\pi r_2)$, with respectively $D = 10^{-6}$ and $D = 10^{-2}$, while $r_1, r_2$ are random numbers chosen in the interval $[0, 1]$. Figures (b) and (d) show the intensity of forcing, as shown in the right hand of Eq. (3), $f \cos[\Omega t + \sigma B(t) + \gamma]$, with respectively $D = (\sigma/2)^2 = 10^{-6}$ and $D = 10^{-2}$, $f = 0.04, \Omega = 1$ and $\gamma$ is a random number chosen in the interval $[0, 2\pi)$.

### 3.2. Computing the appearance of the Wada property in the exit basin in function of the external excitation and the dissipation

As we discussed earlier, the capacity of predicting the behavior of a system is crucial in science and engineering, so when some unpredictabilities show up in the system, their control becomes important. In this section, we analyze by numerical simulations, how the damping can help us to recompose the basins and reduce the merging of the basins. To achieve this goal, we calculate the basins keeping constant the initial energy, $E = 0.25$, and changing both the dissipative parameter and the noisy excitation, evaluating every single case. When the external forcing and the damping are changed, we have seen that it is possible to discern when the structure of the basins looses the coherence and the boundaries start to mix. For bigger values of dissipation and a lower value of the forcing, the basin structures are connected and compact, as shown in Fig. 6(a). While when we decrease the damping and increase the forcing the same boundaries start to intermingle as we see in Fig. 6(b). Now, in order to distinguish the cases between Wada and non-Wada a formal method is needed and it is provided by the theorem of Kennedy and Yorke [25]. It states that, if P is a periodic point on the basin boundary, the following two conditions are satisfied: (1) its unstable manifold intersects every basin (main condition), and (2) this is the only periodic point accessible from the basin of interest, then the basins have the Wada property. This last property, for the Hénon–Heiles system, has been shown in Ref. [26]. Concerning to the main property, we show in Fig. 6(c) the unstable manifold of the periodic point $P = (1.02461, 0)$, representation of a Lyapunov orbit on the phase space $(y, \dot{y})$, and it intersects all the basins, verifying the conditions of the Kennedy–Yorke theorem. On the other hand, Fig. 6(a) represents the basins of the system for an amount of dissipation and forcing from the non-Wada region.

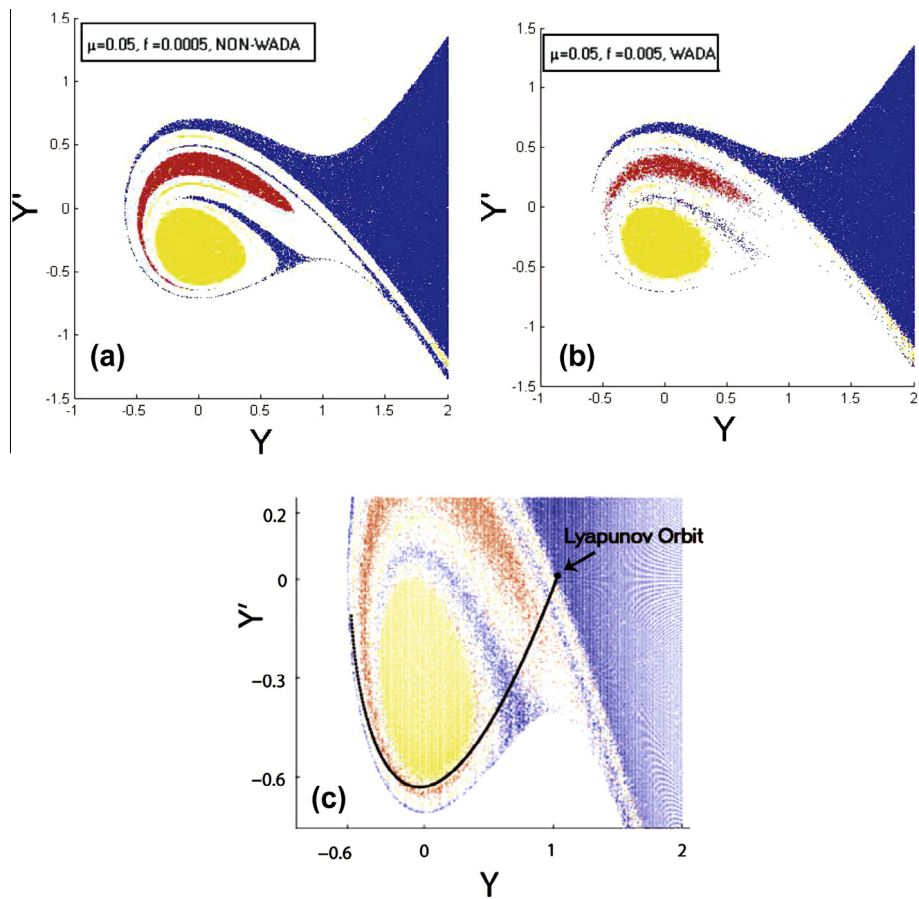

**Fig. 6.** (a) The figure represents respectively, for $E = 0.25$, the basins of the system for an amount of forcing $f = 0.0005$ and damping $\mu = 0.05$, for which the boundaries are coherent. (b) The figure plots respectively, $E = 0.25$, the basins of the system for an amount of forcing $f = 0.005$ and damping $\mu = 0.05$, for which the boundaries are mixed. Here the influence of a bigger external forcing can be observed, making the basins more intermingled than in Fig. 6(a). (c) The figure shows the unstable manifold, the black curve, of the Lyapunov orbit drawn on a zoom of the basins of Fig. 6(b). It is possible to see that the unstable manifold intersects all the basins, so the Wada property is satisfied.

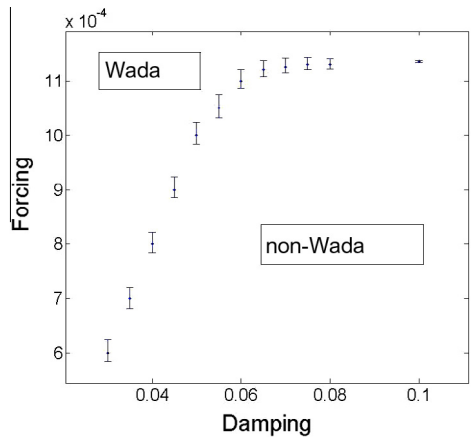

**Fig. 7.** This figure represents, for an energy value $E = 0.25$, the points of the damping-forcing plane for which the Wada property starts to appear in the exit boundaries. The points limit two regions: above them we have the region where Wada basins appear and below the region where we can not find Wada basins. In the non-Wada region, we can predict the evolution of the system while in the Wada region the basin topology is very complicated and the evolution of the system is quite difficult to figure out.

Nevertheless, there is a minimum value of the dissipation for which the basins are not intermingled. Below this value, the basins become Wada even without an external forcing. For this case, the unpredictability of the system increases and the prediction of its evolution becomes impossible. Starting from this point, we increased the dissipation and the excitation to find the limit in which Wada basins appear as shown in Fig. 7.

Thus, two regions appear: the Wada region above the points and the non-Wada region below them. In the non-Wada region, we can predict the evolution of the system while in the Wada region the basin topology is very complicated and the evolution of the system is quite difficult to figure out. The figure also shows on the top right a kind of plateau, as a consequence of a quasi-equilibrium of the external excitation with the damping.

In Fig. 8(a) and (b) we plot the escape time for $y = 0$, $\mu = 0.06$ and different shooting angles. The difference between the figures is the intensity of the external forcing $f$, the first one belonging to the Wada region with a forcing value of $f = 5 \times 10^{-3}$, while the second one to the non-Wada region, with $f = 5 \times 10^{-4}$.

It is possible to see the difference between the mean escape time, where Fig. 8(a) shows a smaller mean escape time than Fig. 8(b). Therefore, as we have thought, Wada basins are related with a smaller mean escape time.

After having computed other escape times in the Wada and non-Wada regions, below and above the points shown in Fig. 7, we have found a similar trend. In fact, we obtain more or less the same results that we have shown in Fig. 8(a) and (b) as the forcing increases.

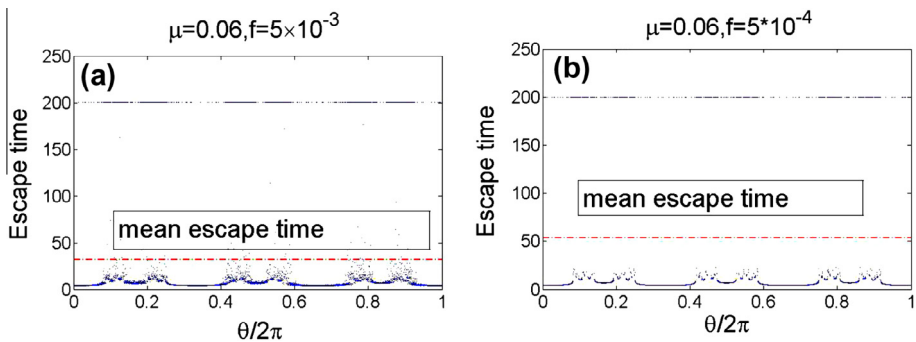

**Fig. 8.** (a) and (b) Both figures represent the escape time for an $E = 0.25$, with dissipation $\mu = 0.06$, versus the shooting angle $\theta/2\pi$. In figure (a) forcing amplitude is $f = 5 \times 10^{-3}$ and in figure (b) forcing amplitude is $f = 5 \times 10^{-4}$. The amplitude of the noisy excitation helps the particles to escape from the scattering region as observed in panel (a). Note that the mean escape time (dash-dot line) is smaller in panel (a) than in panel (b).

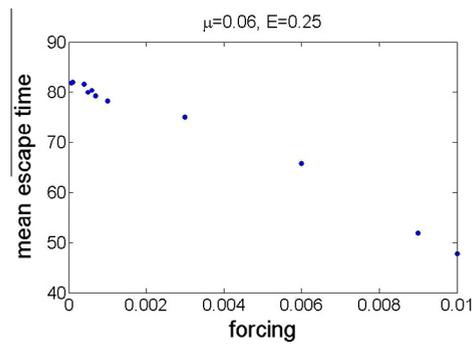

**Fig. 9.** The figure shows the evolution of the mean escape time for a constant amount of dissipation, $\mu = 0.06$, and energy, $E = 0.25$, with respect to the variation of the external excitation. As it can be observed, the higher the forcing amplitude implies the lower mean escape time, since the particles escape faster from the scattering region insofar the forcing amplitude $f$ increases.

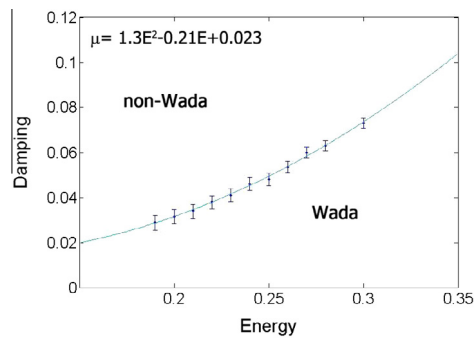

**Fig. 10.** The figure shows the relation between the minimum value of the damping and the initial energy for which Wada basins appear, where there is no external excitation, $f = 0$. We can observe a quadratic fit, $\mu \sim E^2$, which separates both regions, non-Wada and Wada.

We have also studied the escape times varying the amplitude of the external excitation for a fixed amount of dissipation, $\mu = 0.06$, and for a value of the energy $E = 0.25$, as depicted in Fig. 9. Here, we can see that increasing the external forcing for values beyond 0.001, the mean escape time decreases strongly. As a consequence, the higher forcing amplitude implies the lower mean escape times since the particles escape faster from the scattering region insofar the forcing amplitude $f$ increases. Actually, beyond that value the basins start to intermingle faster, so that the unpredictability increases, i.e., the dissipation-induced basins start to scatter. While the forcing helps the trajectories to leave the scattering region, the bounded noise produces a mixing in the basins and the mean escape times decrease.

We have discussed earlier that it is possible to find a minimum value for the dissipation, for which the basins do not show the Wada property. So we have decided to investigate the relation between this minimum and the energy. We consider here a large range of initial energy values for which the motions are unbounded, between 0.19 and 0.3, and we use a forcing $f = 0$ in order to analyze the relationship between the damping and the initial energy.

We show this relationship in Fig. 10, where we observe that as the initial energy increases, the minimum value of the damping also increases. This polynomial-like curve of the *uncertainty boundary U*, has a quadratic fit $\mu \sim E^2$, and its mathematical expression is given by $\mu = 1.3E^2 - 0.21E + 0.023$. A possible explanation of this phenomenon lies in the integration equations in which the dissipation is a factor of the velocity, the equation of which is $v_i = \sqrt{2E - y_i^2 - 2/3 y_i^3}$. However, the effect of the dissipation on the system has to be the same for all the energies, that is to crash the connectedness and compactness of the basins in order to induce the Wada property even without the external excitation, $f = 0$. Therefore, if in Eq. (2) the forcing is equal to zero, the damping value has to grow up with the energy as a factor of the velocity in a polynomial way, through Eq. (4) as depicted in Fig. 10.

Even if, by looking at the tendency curve, it is possible to observe that it matches very well with the data, we report some statistical evidence of the goodness of this fit, like the correlation $R^2 = 0.997$ and the root mean square error $rmse = 0.00088$.

## 4. Conclusions

We have studied in detail the dynamics of the randomly driven and dissipative Hénon–Heiles Hamiltonian. We consider the system subjected to dissipation and a random driven forcing, in the range of initial energy values higher than the escape energy, $E_e = 0.16$, where therefore there exists three exits for the trajectories to leave the scattering region. In order to analyze the relationship between the external forcing, the dissipation and the associated uncertainty, we have considered trajectories inside the scattering region under different conditions of the perturbations and analyzed the way they escape outside from the scattering region. This study permitted us to compute the exit basins. We have seen that for different values of forcing and damping the basins could present the Wada property or not, which is directly related with the unpredictability of the system. We have studied, via numerical simulations, for what amount of external forcing the basins start to intermingle, enhancing the unpredictability of the system. Then, we have studied for what amount of dissipation the basins loose the Wada property, becoming more predictable, and repeated everything for different values of the initial energy. We think our results are useful to gain a better understanding on the possibility to control the unpredictability in this kind of systems, through the use of energy dissipation. We found that it is possible to find a minimum of the dissipation for which the basins are not Wada, but still compact. We have computed this minimum value for different energies and we have found a polynomial relation between the energy and the dissipation. Moreover, we have calculated, for a fixed initial energy, the pair of damping and forcing values, for which the basins start to show the Wada property. This analysis can be useful to know, in a system that presents Wada basins in phase space, where they appear in order to understand better where the system presents more unpredictability and ways to control it.

## Acknowledgment

We acknowledge financial support by the Spanish Ministry of Science and Innovation under Project No. FIS2009-09898.